%% file: Vinet_P5O-6.tex
\documentclass[graybox]{svmult}

\usepackage{mathptmx}       % selects Times Roman as basic font
\usepackage{helvet}         % selects Helvetica as sans-serif font
\usepackage{courier}        % selects Courier as typewriter font
\usepackage{type1cm}        % activate if the above 3 fonts are
\usepackage{makeidx}         % allows index generation
\usepackage{graphicx}        % standard LaTeX graphics tool
\usepackage{multicol}        % used for the two-column index
\usepackage[bottom]{footmisc}% places footnotes at page bottom
\usepackage{amsmath, amssymb}
\usepackage{units}

\newcommand{\ket}[1]{|#1\rangle\,}
\newcommand{\bra}[1]{\langle #1|\,}

\makeindex             % used for the subject index
\begin{document}

% Please name your Latex file as lastname_session.tex
% where session should be included as in the following examples:
% - lastname_M2O.tex for oral talk in M2,
% - lastname_P5P.tex for poster in P5,
% - lastname_PL.tex for plenary talks

\title*{Exact State Revival in a Spin Chain with Next-To-Nearest Neighbour Interactions}
\author{Matthias Christandl$^\dagger$, Luc Vinet$^\star$ and Alexei Zhedanov$^\ddagger$}
\institute{$^\dagger$Matthias Christandl \at Department of Mathematical Sciences, University of Copenhagen, Universitetspanken 5, 2100 Copenhagen, Denmark \\ \email{christandl@math.ku.dk}
\and $^\star$Luc Vinet \at Centre de Recherches Math\'{e}matiques, Universit\'{e} de Montr\'{e}al, C.P. 6128, Succursale Centre-ville, Montr\'{e}al, QC, Canada, H3C 3J7 \\ \email{vinet@crm.umontreal.ca}
\and $^\ddagger$Alexei Zhedanov \at Department of Mathematics, Information School, Renmin University of China, Beijing 100872, China. \\ \email{zhedanov@yahoo.com}}
\maketitle

\abstract{An extension with next-to-nearest neighbour interactions of the simplest XX spin chain with perfect state transfer (PST) is presented. The conditions for PST and entanglement generation (balanced fractional revival) can be obtained exactly and are discussed.}

\section{Introduction}
\label{sec:1}
Certain spin chains have been known to model advantageously devices that effect perfectly the transfer of quantum states between locations [1, 2, 3]. Calling upon their dynamics to realize the transport has the merit of minimizing the need for external interventions and of protecting coherence. Analytic models have been found for which the ocurrence of this perfect state transfer (PST) is demonstrated from an exact analysis. As a rule the couplings between spins must be non-uniform. The simplest such spin chain is of the XX type with parabolic couplings only between its nearest neighbours [4]. It is referred to as the Krawtchouk model in view of the family of orthogonal polynomials that emerge in its description. This model is proving quite useful not only as a paradigm example but also as a test bed for experimentalists. As it turns out, these spin models have a translation in terms of arrays of optical waveguides in view of the mathematical equivalence of the single excitation dynamics of spin chains with the coupled mode theory of optical lattices. Recent experimental implementations [5, 6] have in fact been carried in this framework using the Kratwchouk model. Restricting to nearest-neighbour (NN) interactions is obviously an approximation in this context and it becomes relevant to examine, exactly if possible, the situation beyond this restriction. It is with this perspective that we present in Sect. 2 an analytic extension of the NN Krawtchouk model that includes next-to-nearest neighbour (NNN) couplings. The conditions for PST along that chain can again be found exactly and will be given in Sect. 3.

There is another phenomenon of importance for quantum information that can be realized in spin chains, namely entanglement generation. This is obtained by end-to-end balanced fractional revival whereby a wavepacket initially at one end is reproduced simultaneously (with half the intensity) at both ends. The NN Krawtchouk model does not exhibit this effect but it proves possible when NNN interactions are included. This will be covered in Sect. 4. A summary and remarks on experiments that this analysis suggests will form the concluding section.

\section{The model}
\label{sec:2}
We shall condier a spin chain with the following Hamiltonian of type XX on $\left(\mathbb{C}^2\right)^{\otimes(N+1)}$ where each of the $(N+1)$ spins interacts with its nearest and next-to-nearest neighbours on the left and on the right:
\begin{equation}
H = \frac{1}{2}\sum_{\ell=0}^{N-1}\left[ J_{\ell+1}^{(1)}\left( \sigma_{\ell}^{x}\sigma_{\ell+1}^{x}+\sigma_{\ell}^{y}\sigma_{\ell+1}^{y} \right) + J_{\ell+2}^{(2)}\left( \sigma_{\ell}^{x}\sigma_{\ell+2}^{x}+\sigma_{\ell}^{y}\sigma_{\ell+2}^{y} \right)\right]
+ \frac{1}{2}\sum_{\ell=0}^{N} B_\ell\left( \sigma_\ell^z +1\right).
\end{equation}
As usual, $\sigma_\ell^x$, $\sigma_\ell^y$, $\sigma_\ell^z$ stand for the Pauli matrices with the index $\ell$ indicating on which of the $\mathbb{C}^2$ factors they act. The nearest-neighbours couplings are taken to be the same as those of the Krawtchouk model : $J_n^{(1)}= \beta J_n$ with $J_n = \frac{1}{2}\sqrt{n(N-n+1)}$ and $\beta$ a parameter. The next-to-nearest neighbour couplings are given by $J_n^{(2)}=\alpha J_{n-1}J_n$ with $\alpha$ another parameter and the local magnetic fields are $B_n = \alpha\left(J_n^2+J_{n+1}^2\right)$.
Note that when $\alpha=0$, the NN Krawtchouk model with no magnetic fields is recovered. Owing to rotational symmetry about the $z$-axis, $H$ preserves the number of spins that are up over the chain, i.e. the number of eigenstates of $\sigma_\ell^z$ with eigenvalue $+1$. In the following, we shall only need to consider chain states that have a single spin up. A natural basis for that subspace is given by the vectors $\ket{n}=(0,0,\cdots,0,1,0,\cdots,0)^\intercal~,~ n=0,\dots,N$, with the only $1$ in the n$^{\text{th}}$ position corresponding to the only spin up at the n$^{\text{th}}$ site. The action of $H$ on those states is given by $H\ket{n}=J_{n+2}^{(2)}\ket{n+2}+J_{n+1}^{(1)}\ket{n+1}+B_n\ket{n}+J_{n}^{(1)}\ket{n-1}+J_{n}^{(2)}\ket{n-2}$. Now consider the operator $J$ that acts as follows on the vectors $\ket{n}$ : $J\ket{n}=J_{n+1}\ket{n+1}+J_{n}\ket{n-1}$. It follows that $J^2\ket{n}=J_{n+1}J_{n+2}\ket{n+2}+\left(J_{n+1}^2+J_n^2\right)\ket{n}+J_nJ_{n-1}\ket{n-2}$. 
We thus observe that
\begin{equation}
\label{eq:hamilt_n}
H\ket{n}=\left(\alpha J^2+\beta J\right)\ket{n}~.
\end{equation}
Let $\ket{x_s}$ be the eigenstates of $J$ with eigenvalues $x_s$ : $J\ket{x_s} =x_s\ket{x_s}$. In view of Eq.~\eqref{eq:hamilt_n}, these will be eigenstates of $H$ with eigenvalues $E_s = \alpha x_s^2+\beta x_s$. As it turns out, the eigenvalues and eigenvectors of $J$ can be obtained from angular momentum theory. Let $L_z$ and $L_\pm$ be the $\mathfrak{su}(2)$ generators represented in the standard fashion by
\begin{align}
\begin{aligned}
L_z\ket{\ell,m} =m\ket{\ell,m}~~,~~
L\pm\ket{\ell,m}=\sqrt{(\ell\mp m)(\ell\pm m+1)}\ket{\ell,m\pm1}
\end{aligned}
\end{align}
on the usual angular momentum states $\ket{\ell,m}$, $-\ell\leq m\leq\ell$. \\
Identify $\ket{\ell,m} =\ket{\frac{N}{2},n-\frac{N}{2}}\equiv\ket{n}~,\ n=0,1,\cdots,N$. Then 
$L_x\ket{n} =\frac{1}{2}\left(L_++L_-\right)\ket{n}=J_{n+1}\ket{n+1}+J_n\ket{n-1}$ and the action of $L_x$ is seen to be that of $J$. Since $L_x=e^{-i\tfrac{\pi}{2}L_y}L_ze^{i\tfrac{\pi}{2}L_y}$, the spectrum of $L_x=J$ is the same as the spectrum of $L_z$, thus $x_s=s-\frac{N}{2}$.
Now consider the expansion of the eigenstates $\ket{x_s}$ on the vectors of the occupational basis
\begin{equation}
\ket{x_s}=e^{-i\tfrac{\pi}{2}L_y}\ket{s}=\sum_{n=0}^{N}\bra{n}e^{-i\tfrac{\pi}{2}L_y}\ket{s}\ket{n}=\sum_{n=0}^{N}\sqrt{\omega_s}\chi_n(x_s)\ket{n}~.
\end{equation}
At this point, either from the 3-term recurrence relation $J_{n+1} \chi_{n+1}(x)+J_n\chi_{n-1}(x)=x\chi_n(x)$ that follows from $J\ket{x_s}=x_s\ket{x_s}$ or the knowledge of the Wigner $\mathcal{D}$ functions, we find that the expansion coefficients are given by the normalized Krawtchouk polyomials which are defined as follows :
\begin{align}
\chi_n(x)=(-1)^n\sqrt{\binom{N}{n}}\ {}_{2}F_{1}\left({-n,-s} \atop {-N} \middle| 2 \right)
\end{align}
with the hypergeometric series given by
\begin{equation}
{}_{2}F_{1}\left({a,b} \atop {c} \middle| z \right)=\sum_{k=0}^{\infty}\frac{(a)_k\ (b)_k}{(c)_k}\frac{z^k}{k!}
\end{equation}
and $(a)_k=a(a+1)\cdots (a+k-1)$. These polyomials are orthogonal with respect to the binomial distribution : $\omega_s=\frac{N!}{s!(N-s)!} \left(\frac{1}{2}\right)^N$. Since $\bra{n}e^{-i\tfrac{\pi}{2}L_y}\ket{s}=\sqrt{\omega_s}\chi_n(x_s)$ are elements of an orthogonal matrix, we also have the inverse expansion
\begin{equation}
\label{eq:expans_n}
\ket{n} =\sum_{s=0}^{N}\sqrt{\omega_s}\chi_n(x_s)\ket{x_s}~.
\end{equation}
For later purposes, observe that when $n=N$ :
\begin{equation}
\label{eq:minus1_Ns}
\chi_N(x_s)=(-1)^N\sum_{k=0}^{N}(-s)_k\frac{2^k}{k!}=(-1)^N\sum_{k=0}^{N}\frac{s!}{(s-k)!}\frac{(-2)^k}{k!}=(-1)^{N+s}~.
\end{equation}

\section{Perfect State Transfer}
\label{sec:2}
Let us examine the conditions for PST, that is the transfer with probability one, after time $T$, of a spin up from one end of the chain to the other. This will happen if
\begin{equation}
\label{eq:condition_1}
e^{-iTH}\ket{0}=e^{i\phi}\ket{N}
\end{equation}
where $\phi$ is some phase. In order to analyze this condition, use the expansion in Eq.~\eqref{eq:expans_n} in terms of the eigenstates of $H$ with eigenvalues $E_s = \alpha x_s^2+\beta x_s$ to find that Eq.~\eqref{eq:condition_1} amounts to $e^{-i\phi} e^{-iTE_s}=\chi_N(x_s)=(-1)^{N+s}$ in view of Eq.~\eqref{eq:minus1_Ns}. This last equation can be rewritten as follows in terms of the exponents :
\begin{equation}
\label{eq:condition_TEs}
T E_s=-\phi+\pi\left(N+s+2L_s\right) ~,\quad s=0,1,\dots,N
\end{equation}
where $L_s$ are arbitrary integers that may depend on $s$. Let us consider first the NN model, with $\alpha=0$, and verify that PST occurs. In this case $E_s=\beta\left(s-\tfrac{N}{2}\right)$ and one has
\begin{equation}
\label{eq:Tbeta_condition}
T \beta\left(s-\tfrac{N}{2}\right)=-\phi+\pi\left(N+s+2L_s\right)~.
\end{equation}
This shows that the integer numbers $L_s$ must depend linearly on $s$ and take the form $L_s=\ell s+m$ with $\ell$ and $m$ integers. With $\phi$ appropriately chosen to take care of the constant terms, Eq.~\eqref{eq:Tbeta_condition} reveals that PST will be achieved at times $T$ given by $T=\tfrac{\pi}{\beta}(2\ell+1)~,\quad \ell=0,1,\dots $ with the minimal time for PST in the NN model being $T=\tfrac{\pi}{\beta}$.

Can PST be maintained in the presence of NNN interactions? The answer is in the affirmative provided certains conditions are verified by the parameters $\alpha$ and $\beta$. When $\alpha\neq0$, the eigenvalues $E_s$ of $H$ are given by $E_s=\alpha\left(s-\tfrac{N}{2}\right)^2+\beta\left(s-\tfrac{N}{2}\right)$ and condition \eqref{eq:condition_TEs} reads
\begin{equation}
T\left[\alpha\left(s-\tfrac{N}{2}\right)^2+\beta\left(s-\tfrac{N}{2}\right)\right]=-\phi+\pi N+\pi s+2\pi L_s~.
\end{equation}
Although more involved, the analysis of this equation proceeds in a way analogous to that of Eq.~\eqref{eq:Tbeta_condition}. The reader will find the details in [7]. The upshot is the following. In distinction to the NN model, PST does not always occur. It will happen in the model with NNN interactions if $\tfrac{\alpha}{\beta}$ is rational, in other words if $\tfrac{\alpha}{\beta}=\tfrac{p}{q}$, with $p$ and $q$ co-prime integers. The minimal PST time is $T=\tfrac{\pi}{\beta}q$. Moreover if $p$ is odd, $q$ and $N$ must be either both odd or both even.

\section{Fractional Revival}
\label{sec:4}
We discuss next the possibility of observing fractional revival (FR) at the two ends of the chain. This FR phenomenon will occur after time $\tau$ if
\begin{equation}
\label{eq:FR_condition}
e^{-iH\tau} \ket{0}=\mu\ket{0}+\nu\ket{N}
\end{equation}
with $|\mu|^2+|\nu|^2=1$. Note that PST is a special case of FR with $\mu=0$ ($|\nu|=1$). Furthermore, it is readily recognized that when $|\mu|=|\nu|=\tfrac{1}{\sqrt{2}}$, the state obtained at time $\tau$ is maximally entangled as a balanced coherent sum of $\ket{0}=\ket{\uparrow\downarrow\downarrow\cdots\downarrow}$ and $\ket{N}=\ket{\downarrow\downarrow\cdots\downarrow\uparrow}$. Now upon using expansion~\eqref{eq:expans_n}, condition~\eqref{eq:FR_condition} is translated into $e^{-i\tau E_s}=e^{i\phi}\left(\mu'+\nu'(-1)^{N+s}\right)$,\ $\mu=e^{i\phi}\mu'$, $\nu=e^{i\phi}\nu'$ and $\mu'$ is chosen real without loss of generality. Taking the modulus on both sides we see that $Re(\mu'\nu')=0$. Given that $\mu'$ is real, $\nu'$ must thus be imaginary. We shall write $\mu'=\cos\theta$, $\nu'=i\sin\theta$ which makes the FR condition become
\begin{equation}
\label{eq:FR_condition2}
e^{-iE_s\tau}=e^{i\phi}\left(\cos\theta+i(-1)^{N+s}\sin\theta\right)~.
\end{equation}
In this parametrization, up to integer multiples of $\pi$, $\theta=\tfrac{\pi}{2}$ corresponds to PST. The conditions for FR at two sites in NN spin chains of type XX have been thoroughly analyzed in [8]. Let us first examine here if FR can be found in the NN Krawtchouk model. For $E_s=\beta\left(s-\tfrac{N}{2}\right)$,~\eqref{eq:FR_condition2} splits into the following two Eqs. according to the parity of $s$ :
\begin{align}
\label{eq:two_eq}
\begin{aligned}
&\beta\tau\left(2s+j-\tfrac{N}{2}\right) =-\phi-(-1)^{N+j} \theta+2\pi L_s^{(j)}~,
\end{aligned}
\end{align}
where $L_s^{(j)}$, $j=0,1$ are two independent sequences of integers that must be of the form $L_s^{(j)}=\gamma_js+\delta_j$, with $\gamma_j$ and $\delta_j$ integers. It follows from~\eqref{eq:two_eq} that $\gamma_0=\gamma_1=1,2,\dots$ and that $\tau=\pi\tfrac{\gamma_0}{\beta}$. Moreover, apart from a relation determining the phase $\phi$ in terms of the parameters, one finds that $\theta=(-1)^N\left[\tfrac{\gamma_0}{2}+(\delta_0-\delta_1)\right]\pi$. Therefore, up to sign and integer multiples of $\pi$, $\theta$ can only take the values $0$ and $\tfrac{\pi}{2}$. This means that only PST and perfect return are possible. We thus reach the conclusion that FR at two sites cannot happen in the NN Krawtchouk model.
Let us now turn to the NNN extension. In this case, the FR condition~\eqref{eq:FR_condition2} yields relations analogous to~\eqref{eq:two_eq} with the l.h.s replaced by $\left[\alpha\left(2s+j-\tfrac{N}{2}\right)^2+\beta\left(2s+j-\tfrac{N}{2}\right)\right]\tau$ and the sequences of integers $L_s^{(j)}$ having instead a quadratic form : $L_s^{(j)}=\xi_j s^2+\eta_j s+\zeta_j$, $j=0,1$ where for each $j$, independently, $\xi_j$ and $\eta_j$ can be simultaneously integer or half-integer while $\zeta_j$ is integer. Once again, we refer the reader to [7] for the detailed analysis of what these equations entail. The findings are as follows. FR can happen in NNN spin chains that have $\tfrac{\alpha}{\beta}=\tfrac{p}{q}$ with $p$ and $q$ co-prime integers and $p$ odd; again, $q$ and $N$ must have the same parity. When these conditions are met $\theta\simeq\tfrac{\pi}{4}$, entanglement generation or balanced FR will be realized and its first occurence will be observed at time $\tau=q\tfrac{\pi}{2\beta}$. 

The picture with respect to FR is thus as follows. While it does not occur in the NN Krawtchouk spin chain, the presence of additional NNN interactions allows this phenomenon to take place under the circumstances that we have spelled out. However, the only form of FR at sites $0$ and $N$ that can be realized is of the balanced type which corresponds to the generation of maximally entangled state.

\section{Conclusion}
\label{sec:5}
Summing up, we have provided an analytic model with NNN interactions that extends the simplest XX spin chain with PST, namely the NN Krawtchouk model. This extended model involves two parameters $\alpha$ and $\beta$. The NN model is recovered when $\alpha=0$. When $\alpha\neq0$, for PST to occur we must have $\tfrac{\alpha}{\beta}=\tfrac{p}{q}$ where $p$ and $q$ are co-prime integers. If FR is to happen, it can only be of the balanced type and $p$ must be odd and in that case $N$ must be of the same parity as $q$.

It would now be quite interesting to obtain an experimental validation of these results. Discussions are underway regarding the design of an optical array in which entanglement generation would be observed as per the predictions and specifications of the analysis that we have described here.

\begin{acknowledgement}
This paper was completed during a stay of LV at the School of Mathematical Sciences of the Shanghai Jia Tong University as Chair Visiting Professor. M.C. acknowledges financial support from the European Research Council (ERC Grant Agreement no 337603), the Danish Council for Independant Research (Sapere Aude) and the Swiss National Science Foundation (project no PP00P2-150734). The research of L.V. is supported by the Natural Sciences and Engineering Council (NSERC) of Canada.
\end{acknowledgement}
\input{referenc}

\end{document}

%% file: referenc.tex
%%%%%%%%%%%%%%%%%%%%%%%% referenc.tex %%%%%%%%%%%%%%%%%%%%%%%%%%%%%%
% sample references
% Use this file as a template for your own input.
%
%%%%%%%%%%%%%%%%%%%%%%%% Springer-Verlag %%%%%%%%%%%%%%%%%%%%%%%%%%
%
% BibTeX users please use
% \bibliographystyle{}
% \bibliography{}
%
% \biblstarthook{References should be \textit{cited} in the text by number.\footnote{Make sure that all references from the list are cited in the text. Those not cited should be moved to a separate \textit{Further Reading} section or chapter.} The reference list should ideally be \textit{sorted} in alphabetical order -- even if reference numbers are used for the their citation in the text. If there are several works by the same author, the following order should be used: 
% \begin{itemize}
% \item all works by the author alone, ordered chronologically by year of publication
% \item all works by the author with a coauthor, ordered alphabetically by coauthor
% \item all works by the author with several coauthors, ordered chronologically by year of publication.
% \end{itemize}
% %
% The style for references is depicted here
% }